\documentclass[smallextended]{svjour3} 

\smartqed

\usepackage{graphicx}
\usepackage{mathptmx}

\RequirePackage{fix-cm}

\journalname{GeNeDis}

\title{An Architecture For Cooperative Mobile Health Applications}
\titlerunning{An Architecture For Cooperative Mobile Health Applications}
\author{%
    Georgios Drakopoulos  \and
    Phivos Mylonas  \and
    Spyros Sioutas
}%
\authorrunning{Drakopoulos et al.}
\institute{%
    Georgios Drakopoulos and Phivos Mylonas \at
    Department of Informatics, Ionian University, Greece\\
    \email{\{c16drak, fmylonas\}@ionio.gr}
\and
    Spyros Sioutas \at
    Computer Engineering and Informatics Department, University of Patras, Greece\\
    \email{sioutas@ceid.upatras.gr}
}%

\date{Received: date / Accepted: date}

\usepackage{amssymb, relsize}
\usepackage{listings, algorithmic, algorithm}
\usepackage{natbib}

\RequirePackage{amsmath, amssymb, relsize}
\DeclareMathOperator{\bmopmean}{E}
\DeclareMathOperator{\bmopvar}{Var}

\newcommand{\bmdef}{{\buildrel{{\mathsmaller{\triangle}}}\over{=}}}

\newcommand{\bmset}[1]{{\left\lbrace{#1}\right\rbrace}}
\newcommand{\bmcardinality}[1]{{\left\lvert{#1}\right\rvert}}

\newcommand{\bmparen}[1]{{\left({#1}\right)}}

\newcommand{\bmmean}[1]{{\bmopmean}\left[{#1}\right]}
\newcommand{\bmvar}[1]{{\bmopvar}\left[#1\right]}

\def\bmwork{chapter}

\begin{document}

\maketitle

\begin{abstract}
Mobile health applications are steadily gaining momentum in the modern world given the omnipresence of various mobile or WiFi connections. Given that the bandwidth of these connections increases over time, especially in conjunction with advanced modulation and error-correction codes, whereas the latency drops, the cooperation between mobile applications becomes gradually easier. This translates to reduced computational burden and heat dissipation for each isolated device at the expense of increased privacy risks. This \bmwork{} presents a configurable and scalable edge computing architecture for cooperative digital health mobile applications.
\end{abstract}

\section{Introduction}\label{sec:intro}

\paragraph{}  Mobile smart applications for monitoring human health or processing health-related data are increasing lately at an almost geometric rate. This can be attributed to a combination of social and technolgical factors. The accumulated recent multidisciplinary research on biosignals and the quest for improved biomarkers bore fruits in the form of advanced bisignal processing algorithms. Smartphone applications are progressively becoming popular in all age groups, albeit with a different rate for each such group and, moreover, mobile subscribers tend to be more willing to provide sensitive health data such are heart beat rate, blood pressure, or eye sight status to applications for processing. Thus, not only technological but also financial factors favor the development of digital health applications. 

\paragraph{}  The primary contribution of this \bmwork{} is a set of guidelines towards a cross-layer cooperative architecture for mobile health applications. The principal motivation behind them are increased parallelism, and consequently lower turnaround or wallclock time, additional redundancy, which translates to higher reliability, and lower energy consumption. All these factors are critical for mobile health applications.

\paragraph{}  The remaining of this \bmwork{} is structured as follows. Section \ref{sec:work} briefly summarizes the recent scientific literature in the fields of edge computing, mobile applications, mobile services, and digital health applications. Section \ref{sec:arch} presents the proposed architecture. Finally, section \ref{sec:more} recapitulates the main points of this \bmwork. The notation of this \bmwork{} is shown at table \ref{tab:symbols}.

\begin{table}[ht]
\caption{\label{tab:symbols}Notation of this \bmwork.}
\begin{tabular}{ll}
    \hline\noalign{\smallskip}
    Symbol     &  Meaning  \\
    \noalign{\smallskip}\hline\noalign{\smallskip}
    $\bmdef$                                                            &  Definition or equality by definition\\
    $\bmset{s_1, \ldots, s_n}$                                          &  Set comprising of elements $s_1, \ldots, s_n$\\
    $\bmcardinality{S}$ or $\bmcardinality{\bmset{s_1, \ldots, s_n}}$   &  Cardinality of set $S$\\
    $\bmmean{X}$                                                        &  Mean value of random variable $X$\\
    $\bmvar{X}$                                                         &  Variance of random variable $X$\\
    $\gamma_1$                                                          &  Skewness coefficient\\
    \noalign{\smallskip}\hline
\end{tabular}
\end{table}

\section{Previous Work}\label{sec:work}

\paragraph{}  Mobile health applications cover a broad spectrum of cases as listed for instance in \cite{sunyaev:2014} or in \cite{fox:2010}. These include pregnancy as described in \cite{banerjee:2013}, heart beat as mentioned in \cite{steinhubl:2013}, and blood pressure as stated in \cite{logan:2007}. Using mobile health applictions results from increased awareness of the digital health potential as \cite{rich:2014} claims. A major driver for the latter is the formation of thematically related communities in online social media as stated in \cite{ba:2013}. Another factor accounting for the popularity as well as for the ease of health applications is gamification as found in \cite{lupton:2013} and \cite{pagoto:2013}, namely the business methodologies relying on gaming elements as their names suggest - see for instance \cite{deterding:2011a}, \cite{deterding:2011b}, or \cite{huotari:2012}. Gamification can already be found at the very core of such applications as described in \cite{cugelman:2013}.

\paragraph{}  The processing path of any digital health may take several forms as shown in \cite{serbanati:2011}. For an overview of recent security practices for mobile health applications see \cite{papageorgiou:2018}. Path analysis as in \cite{drakop:2017:webist} play a central role in graph mining in various contexts, for instance in social networks as in \cite{drakop:2017:snam}. Finally, the advent of advanced GPU technologies can lead to more efficient graph algorithms as in \cite{drakop:2018:ictai}.

\paragraph{}  Finally, although it has been only very recently enforced (May 2018), GDPR, the EU directive governing the collection, processing, and sharing of sensitive personal information, seems to be already shaping more transparent conditions the smartphone applications ecosystem is adapting to. In fact, despite the original protests that GDPR may be excessively constraining under certain circumstances described in \cite{charitou:2018}, consumers seem to trust mobile applications which clearly outline their intentions concerning any collected piece of personal information as \cite{bachiri:2018} found out.

\section{Architecture}\label{sec:arch}

\begin{figure*}[ht]
\centering%
\includegraphics[scale=0.7]{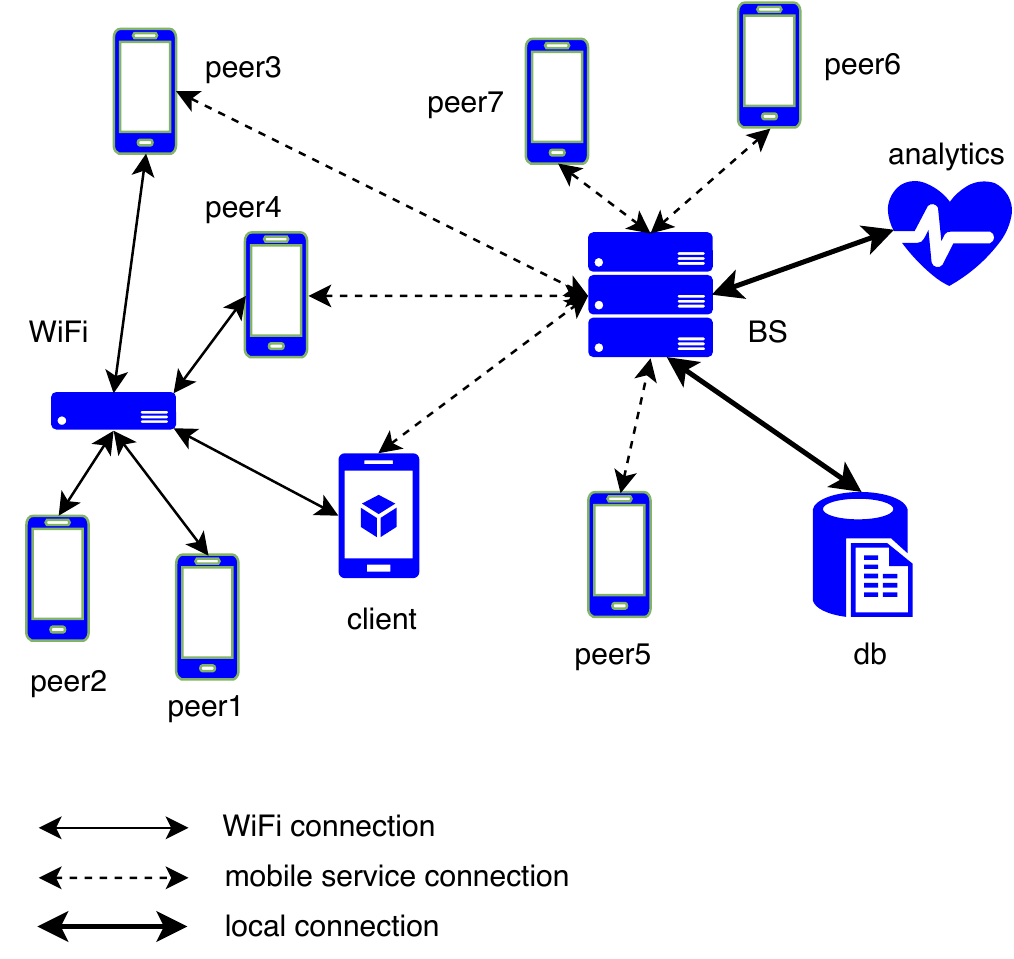}
\caption{\label{fig:arch}Instance of a mobile application surrounded by peers.}
\end{figure*}

\paragraph{}  This section presents and analyzes the proposed cooperative architecture for mobile digital health applications. Figure \ref{fig:arch} visualizes an instance of a mobile health application running on a smartphone and a number of peers which can be reached either by WiFi or by regular mobile services.

\paragraph{}  As with the majority of mobile architectures, the proposed architecture is conceptually best described with graphs, as concepts such as connectivity and community structure can be naturally expressed. To this end, the cell phones, the base stations, and the WiFi access points are represented as vertices, each device category being represented as a different type. Moreover, connections between these are represented as edges, where each edge is also of different type depending on the connection. These can be easily programmed in a graph database like Neo4j.

\paragraph{}  The general constraints that will be the basis for the subsequent analysis are as follows:
\begin{itemize}

\item  Assume that a mobile health application monitoring a biomarker or a biosignal must deliver results every $T_0$ time units, usually seconds. Additionally assuming that the required computation can be split into $n+1$ parts to be distributed to the available $n$ neighbors, then:
\begin{equation}
T_a + T_p + T_s + 2T_c  \leq T_0
\label{eq:time}
\end{equation}
Where $T_a$, $T_p$, $T_s$, and $T_c$ denote respectively the time required for analysis, namely breaking down the computation and assigning each neighbor a task, processing, namely the time of the slowest task, synthesis, namely assemblying the solution of each task to create the general solution, and communication. The latter term counts twice as the data and the task need to be communicated and then the results need to be collected.

\item  In mobile communications is of paramount importance the minimization of the energy dedicated to a single task. In general the relationship between a given task and the energy spent for its accomplishment is unknown. However, given that tasks have a short duration, it is fairly reasonable to assume that the same function $f\bmparen{\cdot}$ links the task and the energy at each neighbor. Then the following inequality should also be satisfied:
\begin{align}
\bmparen{n+1}f\bmparen{T_p} + f\bmparen{T_a} + f\bmparen{T_s} + 2\bmparen{n+1}f\bmparen{T_c} &  \:\leq\: f\bmparen{T_0}  \:\Leftrightarrow\:  \cr
\frac{f\bmparen{T_0} - f\bmparen{T_a} - f\bmparen{T_s}}{f\bmparen{T_p} + 2\bmparen{T_c}}  - 1 &  \:\geq\: n
\label{eq:power}
\end{align}

\end{itemize}

\paragraph{}  Given the fundamental constraints (\ref{eq:time}) and (\ref{eq:power}), let us estimate the key parameter $T_c$, since $T_a$, $T_s$, and $T_p$ depend on the problem and $T_0$ is a constraint. 

\paragraph{}  Let $e_{i,j}$ denote the communication link between vertices $v_i$ and $v_j$ has a given capacity $C_{i,j}$ as well as a propagation delay $\tau_{i,j}$. Then, the number of bits $b_{i,j}$ which can be transmitted over edge $e_{i,j}$ in a time slot of length $\tau_0$ is, assuming the variables are expressed in the proper units: 
\begin{equation}
b_{i,j}  \:=\:
    C_{i,j}\bmparen{\tau_0 - \tau_{i,j}}
\label{eq:bits}
\end{equation}
If the link delay $\tau_{i,j}$ is expressed as a percentage $0 < \rho^{\tau}_{i,j} < 1$ of the time slot $\tau_0$, then:
\begin{equation}
b_{i,j}  \:=\:
    C_{i,j} \tau_0 \bmparen{1-\rho^{\tau}_{i,j}}
\end{equation}
Note that the case $\rho^{\tau}_{i,j} = 0$ represents a near physical impossibility, whereas the case $\rho^{\tau}_{i,j} = 1$ denotes either a useless link or a misconfigured network protocol.

\paragraph{}  In a similar way, if $C_0$ is the maximum capacity, then each $C_{i,j}$ can be expressed as a percentage $0 < \rho^C_{i,j} \leq 1$ of the former. Thus:
\begin{equation}
b_{i,j}  \:=\:
    C_0 \tau_0 \rho^C_{i,j} \bmparen{1 - \rho^{\tau}_{i,j}}  \:=\:
    B_0 \rho^C_{i,j} \bmparen{1 - \rho^{\tau}_{i,j}}
\end{equation}
Note that in this case $\rho^C_{i,j}$ can be $1$, unless $C_0$ is an asymptotically upper limit. Therefore, if for the given task $B_{i,j}$ bits must be transmitted, then the total number of slots for that particular link is:
\begin{equation}
T_{i,j}  \:=\:
    \left\lceil \frac{B_{i,j}}{b_{i,j}}  \right\rceil
\end{equation}

\paragraph{}  At this point, we can estimate $T_c$ as:
\begin{equation}
T_c  \:\bmdef\:
    \bmmean{T_{i,j}}
\end{equation}

\paragraph{}  Furthermore, we can use the distribution of $T_c$ to determine whether a big task should be subdivided to smaller tasks. Assuming $\tau_0$ is constant, then it can be used as a reference point to consider the frequency distribution of $T_{i,j}$, which can be treated as a probability distribution. 

\paragraph{}  For any random variable $X$ is possible to define the skewness coefficient $\gamma_1$ as:
\begin{equation}
\gamma_1 \:\bmdef\:
    \bmmean{\frac{X - \bmmean{X}}{\sqrt{\bmvar{X}}}}  \:=\:
    \frac{\bmmean{X^3} - 3\bmmean{X}\bmvar{X} - \bmmean{X}^3}{\bmvar{X}^{\frac{3}{2}}}
\label{eq:skewness}
\end{equation}
In equation (\ref{eq:skewness}) $\bmmean{X}$ and $\bmvar{X}$ stand for the stochastic mean and variance of $X$ respectively. In actual settings these can be replaced by their sample counterparts and they can be updated as new measurements are collected. In the derivation of the right hand side of (\ref{eq:skewness}) the following properties were used:
\begin{align}
\bmmean{\sum_{k=1}^{n}\alpha_k X_k + \alpha_0}  &  \:=\:
    \sum_{k=1}^n\alpha_k \bmmean{X_k} + \alpha_0  \cr
\bmvar{\alpha_1 X + \alpha_0}  &  \:=\: 
    \alpha_1^2 \bmvar{X}
\label{eq:properties}
\end{align}

\paragraph{}  The skewness sign indicates the shape of the distribution. When $\gamma_1$ is negative, then $X$ takes larger values with higher probability, whereas when $\gamma_1$ is positive, then $X$ is more likely to take lower values. Finally, in the case where $\gamma_1$ is zero, then the distribution of $X$ is symmetric, as for instance in the case of the binomial distribution.

\paragraph{}  Therefore, positive values of the skewness coefficient $\gamma_1$ for the distribution of $T_c$ of the channel delays indicate that it is more likely more time to be available for useful information transmission.

\paragraph{}  The proposed methodology is summarized in algorithm \ref{algo:scheme}.
\begin{algorithm}
\begin{algorithmic}[1]
\caption{\label{algo:scheme}The proposed scheme.}
\REQUIRE  Knowledge of $T_0$, $T_s$, $T_a$, and $T_p$.
\ENSURE  A cooperative computation takes place.
\REPEAT
    \STATE  \textbf{update} estimates for $\bmset{T_{i,j}}$
    \IF{equations (\ref{eq:power}) and (\ref{eq:properties}) are satisfied}
	\STATE  \textbf{break} the problem into tasks
    \ENDIF
    \STATE  \textbf{communicate} tasks
    \STATE  \textbf{compute} tasks
    \STATE  \textbf{collect} results
    \STATE  \textbf{compose} answer
\UNTIL{\TRUE}
\end{algorithmic}
\end{algorithm}

\section{Conclusions}\label{sec:more}

\paragraph{}  This \bmwork{} presents a probabilistic architecture for cooperative computation in mobile health app settings. It relies on a higher order statistical criterion, namely the skewness coefficient of the number of slots which are suitable for communication, in order to estimate whether a computation can be broken into smaller tasks and communicated to neighboring smartphones over WiFi or the ordinary cell network. Once the tasks are complete, the results are collected back at the controling smartphone and an answer is generated using a synthesis of these results.

\paragraph{}  In order to find the hard limits of the proposed architecture and to assess its performance under various operational scenaria, a number of simulations must be run in addition to theoretical probabilistic analysis. Additionally, more conditions should be added to the architecture, for instance what happens when a neighboring smartphone stops working or is moved out of range. Moreover, conditions for duplicating certain critical computation must also be created.

\begin{acknowledgements}
This \bmwork{} is part of Tensor 451, a long term research initiative whose primary objective is the development of novel, scalable, numerically stable, and interpretable tensor analytics.
\end{acknowledgements}

\bibliographystyle{spbasic}
\bibliography{genedis_2018_mobile}

\begin{thebibliography}{20}
\providecommand{\natexlab}[1]{#1}
\providecommand{\url}[1]{{#1}}
\providecommand{\urlprefix}{URL }
\expandafter\ifx\csname urlstyle\endcsname\relax
  \providecommand{\doi}[1]{DOI~\discretionary{}{}{}#1}\else
  \providecommand{\doi}{DOI~\discretionary{}{}{}\begingroup
  \urlstyle{rm}\Url}\fi
\providecommand{\eprint}[2][]{\url{#2}}

\bibitem[{Ba and Wang(2013)}]{ba:2013}
Ba S, Wang L (2013) {D}igital health communities: {T}he effect of their
  motivation mechanisms. {D}ecision {S}upport {S}ystems 55(4):941--947

\bibitem[{Bachiri et~al.(2018)Bachiri, Idri, Fern{\'a}ndez-Alem{\'a}n, and
  Toval}]{bachiri:2018}
Bachiri M, Idri A, Fern{\'a}ndez-Alem{\'a}n JL, Toval A (2018) {E}valuating the
  privacy policies of mobile personal health records for pregnancy monitoring.
  {J}ournal of medical systems 42(8):144

\bibitem[{Banerjee et~al.(2013)Banerjee, Chen, Erman, Gopalakrishnan, Lee, and
  Van Der~Merwe}]{banerjee:2013}
Banerjee A, Chen X, Erman J, Gopalakrishnan V, Lee S, Van Der~Merwe J (2013)
  {MOCA}: {A} lightweight mobile cloud offloading architecture. In:
  {P}roceedings of the eighth {ACM} international workshop on Mobility in the
  evolving internet architecture, {ACM}, pp 11--16

\bibitem[{Charitou et~al.(2018)Charitou, Kogias, Polykalas, Patrikakis, and
  Cotoi}]{charitou:2018}
Charitou C, Kogias DG, Polykalas SE, Patrikakis CZ, Cotoi IC (2018) {U}se of
  apps for crime reporting and the {EU} {G}eneral {D}ata {P}rotection
  {R}egulation. {S}ocietal Implications of Community-Oriented Policing and
  Technology pp 55--61

\bibitem[{Cugelman(2013)}]{cugelman:2013}
Cugelman B (2013) {G}amification: {W}hat it is and why it matters to digital
  health behavior change developers. {JMIR} {S}erious {G}ames 1(1)

\bibitem[{Deterding et~al.(2011{\natexlab{a}})Deterding, Dixon, Khaled, and
  Nacke}]{deterding:2011a}
Deterding S, Dixon D, Khaled R, Nacke L (2011{\natexlab{a}}) {F}rom game design
  elements to gamefulness: {D}efining gamification. In: {P}roceedings of the
  15th international academic {M}ind{T}rek conference: {E}nvisioning future
  media environments, {ACM}, pp 9--15

\bibitem[{Deterding et~al.(2011{\natexlab{b}})Deterding, Sicart, Nacke, O'Hara,
  and Dixon}]{deterding:2011b}
Deterding S, Sicart M, Nacke L, O'Hara K, Dixon D (2011{\natexlab{b}})
  {G}amification. using game-design elements in non-gaming contexts. In:
  {CHI}'11 extended abstracts on human factors in computing systems, {ACM}, pp
  2425--2428

\bibitem[{Drakopoulos et~al.(2017)Drakopoulos, Kanavos, Mylonas, and
  Sioutas}]{drakop:2017:snam}
Drakopoulos G, Kanavos A, Mylonas P, Sioutas S (2017) {D}efining and evaluating
  {T}witter influence metrics: {A} higher order approach in {N}eo4j. {SNAM}
  71(1)

\bibitem[{Drakopoulos et~al.(2018)Drakopoulos, Liapakis, Tzimas, and
  Mylonas}]{drakop:2018:ictai}
Drakopoulos G, Liapakis X, Tzimas G, Mylonas P (2018) {A} graph resilience
  metric based on paths: {H}igher order analytics with {GPU}. In: {ICTAI},
  {IEEE}

\bibitem[{Fox and Duggan(2010)}]{fox:2010}
Fox S, Duggan M (2010) {M}obile health 2010. {P}ew Internet and {A}merican Life
  Project {W}ashington, {DC}

\bibitem[{Huotari and Hamari(2012)}]{huotari:2012}
Huotari K, Hamari J (2012) {D}efining gamification: a service marketing
  perspective. In: {P}roceedings of the 16th international academic
  {M}ind{T}rek conference, {ACM}, pp 17--22

\bibitem[{Kanavos et~al.(2017)Kanavos, Drakopoulos, and
  Tsakalidis}]{drakop:2017:webist}
Kanavos A, Drakopoulos G, Tsakalidis A (2017) {G}raph community discovery
  algorithms in {N}eo4j with a regularization-based evaluation metric. In:
  {WEBIST}

\bibitem[{Logan et~al.(2007)}]{logan:2007}
Logan AG, et~al. (2007) {M}obile phone--based remote patient monitoring system
  for management of hypertension in diabetic patients. {A}merican {J}ournal of
  {H}ypertension 20(9):942--948

\bibitem[{Lupton(2013)}]{lupton:2013}
Lupton D (2013) {T}he digitally engaged patient: {S}elf-monitoring and
  self-care in the digital health era. {S}ocial {T}heory and {H}ealth
  11(3):256--270

\bibitem[{Pagoto and Bennett(2013)}]{pagoto:2013}
Pagoto S, Bennett GG (2013) {H}ow behavioral science can advance digital
  health. {T}ranslational behavioral medicine 3(3):271--276

\bibitem[{Papageorgiou et~al.(2018)Papageorgiou, Strigkos, Politou, Alepis,
  Solanas, and Patsakis}]{papageorgiou:2018}
Papageorgiou A, Strigkos M, Politou E, Alepis E, Solanas A, Patsakis C (2018)
  {S}ecurity and privacy analysis of mobile health applications: {T}he alarming
  state of practice. {IEEE} {A}ccess 6:9390--9403

\bibitem[{Rich and Miah(2014)}]{rich:2014}
Rich E, Miah A (2014) {U}nderstanding digital health as public pedagogy: {A}
  critical framework. {S}ocieties 4(2):296--315

\bibitem[{Serbanati et~al.(2011)Serbanati, Ricci, Mercurio, and
  Vasilateanu}]{serbanati:2011}
Serbanati LD, Ricci FL, Mercurio G, Vasilateanu A (2011) {S}teps towards a
  digital health ecosystem. {J}ournal of {B}iomedical {I}nformatics
  44(4):621--636

\bibitem[{Steinhubl et~al.(2013)Steinhubl, Muse, and Topol}]{steinhubl:2013}
Steinhubl SR, Muse ED, Topol EJ (2013) {C}an mobile health technologies
  transform health care? {JAMA} 310(22):2395--2396

\bibitem[{Sunyaev et~al.(2014)Sunyaev, Dehling, Taylor, and
  Mandl}]{sunyaev:2014}
Sunyaev A, Dehling T, Taylor PL, Mandl KD (2014) {A}vailability and quality of
  mobile health app privacy policies. {J}ournal of the {A}merican {M}edical
  {I}nformatics {A}ssociation 22(e1):e28--e33

\end{thebibliography}
\end{document}